\documentclass[preprint, review,12pt]{elsarticle}

\usepackage{graphicx}
\usepackage{amssymb,amsmath}

\usepackage{lineno}
\usepackage{braket}

\usepackage{color}

\journal{Optics Communications}

\begin{document}

\begin{frontmatter}


\title{Single plane minimal tomography of double slit qubits}



\author{Edwin C. Chaparro Sogamoso}
\author{D. Angulo}
\author{K. M. Fonseca-Romero\corref{mycorrespondingauthor}}
\cortext[mycorrespondingauthor]{Corresponding author}
\ead{kmfonsecar@unal.edu.co}

\address{Universidad Nacional de Colombia - Sede Bogot\'a, Facultad de Ciencias,\\  Departamento de F\'isica, Grupo de \'Optica e Informaci\'on Cu\'antica, \\
Carrera 30 Calle 45-03, C.P. 111321, Bogot\'a, Colombia.}

\begin{abstract}
The determination of the density matrix of an ensemble of identically prepared quantum systems by performing a series of measurements, known as quantum tomography, is minimal when the number of outcomes is minimal.
The most accurate minimal quantum tomography of qubits, sometimes called a tetrahedron measurement, corresponds to projections over four states which can be represented on the Bloch sphere as the vertices of a regular tetrahedron.
We investigate whether it is possible to implement the tetrahedron measurement of double slit qubits of light, using measurements performed on a single plane. 
Assuming Gaussian slits and free propagation, we demonstrate that a judicious choice of the detection plane and the double slit geometry allows the implementation of a  tetrahedron measurement.
Finally, we consider possible sets of values which could be used in actual experiments.


\end{abstract}

\begin{keyword}
Optimal quantum tomography \sep Spatial qubits\sep Interference


\end{keyword}

\end{frontmatter}


\section{Introduction}
\label{S:1}

%

Quantum effects can be exploited to process information in different and sometimes more efficient ways \cite{Nielsen2000} than those allowed by classical physics.
For instance, quantum physics promises faster computers  \cite{Divincenzo1995S} and more secure communications \cite{Gisin2002RMP}. 
Different systems \cite{Xiang2013RMP} (photons, atoms, spins, and superconducting and nanomechanical structures) are being used to build quantum devices.
Among the optical implementations of quantum information technologies, the most popular one uses photon polarization as a natural two-level system (qubit) \cite{Kwiat1995PRL}.
Implementations based on the spatial and temporal \cite{Brendel1999PRL} degrees of freedom of light have also been used. 
For example, orbital angular momentum eigenstates have been employed to define (spatial) quantum d-level systems (qudits)  \cite{Allen1992PRA}. Spatial qudits can also be produced when photons are made to pass through an aperture with $d$ pixels  \cite{Sullivan2005PRL} or with $d$ slits \cite{Neves2004PRA,Neves2005PRL}.
The photon transverse position has been used to prepare, measure and control spatial qudit states \cite{Neves2004PRA,Peeters2009PRA}. 
In particular, several methods to estimate the quantum state of slit qubits (and qudits) have been reported \cite{Neves2005PRL,Neves2007PRA,Lima2007JPCS,Lima2008JPB,Taguchi2008PRA,Pimenta2010OE}.

Due to their common mathematical structure, classical waves and quantum mechanics share many physical effects, such as the Gouy geometrical phase \cite{Simon1993}, to mention just one.
Recently, ideas originated in quantum estimation have been applied to classical contexts.
For example, the quantum Fisher information has been used to show that suitable measurements allow the resolution of incoherent sources   separated by distances which violate the Rayleigh criterion \cite{Paur2016,Tsang2016}.
The physical system that we consider in this work, a double-slit qubit, corresponds to the classical Young's interference experiment.
We present a proposal for minimal tomographic reconstruction of double-slit qubits employing only free propagation.
Although our results have been presented in a quantum mechanics language, analogous results hold for the classical two-slit interference setup when we change photon detection by intensity measurement.

A brief account of quantum tomography, section \ref{sec:QT}, provides context to formulate the problem of state estimation for qubits defined by a two-slit setup
(section \ref{sec:Sec2slit}).
In section \ref{sec:Solution}, we demonstrate how the minimal quantum tomography can be performed using measurements on a single plane.
Finally, conclusions are drawn, and additional remarks, concerning possible experimental implementations, are made.

\section{Quantum Tomography}
\label{sec:QT}

Quantum tomography is an a posteriori process that allows a thorough description of the quantum state of an assembly of identically prepared systems,  based on data obtained with measurement apparatuses \cite{Buzek1998AP}.  
The origin of quantum tomography of systems of continuous variables can be traced back to Pauli \cite{Pauli1933}, who considered the problem of the reconstruction of the wavefunction of a spinless quantum particle, given its coordinate and momentum  probability densities \cite{Orlowski1994PRA}.
In general, the probability density and the probability current (not coordinate and momentum probability densities) allow the reconstruction of pure states \cite{Gale1968PR}.
In the case of mixed states, it is possible to reconstruct the Wigner quasiprobability function from the probability distributions along straight lines in phase space \cite{Bertrand1987FP,Vogel1989PRA}.
Experimental state reconstruction in a diversity of quantum systems (including molecular vibrational modes, one-mode  and two-mode states of light, trapped ions, and helium atoms) have been reported \cite{Buzek1998AP,Raymer2004LNP}. 

Stokes \cite{Stokes1851TCPS} arguably is the father of tomographic methods for the reconstruction of systems with finite-dimension Hilbert spaces. 
However, the first systematic approach to state estimation is due to Fano \cite{Fano1957RMP}, who introduced the notion of a quorum, a set of observables sufficient to determine the quantum state of a system.
Any quantum tomography ---spin tomography \cite{DAriano2000PLA, Weigert2000PRL,Hofmann2004PRA}, for example--- can be performed using different quora.
The elements of a quorum are not necessarily associated with observables, but with positive semidefinite operators $\mathbb{P}_m$ which resolve the identity, $\mathbb{I}=\sum_{m=1}^k \mathbb{P}_m$.
This set of operators, collectively known as a positive operator valued measure (POVM), describe generalized measurements.

When the statistics of a POVM can completely determine the quantum state of a system, it is said to be informationally complete (IC).
An IC-POVM must contain at least $d^2$ operators $P_m,$ to be able to estimate the $d^2-1$ reals parameters that uniquely determine the density matrix of a $d$-level system.
Two prominent examples of IC measurements are mutually unbiased  (MU) measurements and symmetric informationally complete (SIC) POVMs.
Bases such that the angles between arbitrary pairs of elements of different bases are all equal are known as MU bases (MUBs) \cite{Schwinger1960PNAS}; a set of $d+1$ MUBs of a system of $d$ levels is informationally complete \cite{Wootters1989AP}.
When the Hilbert-Schmidt inner products between every pair of different operators of an IC-POVM are all equal, this POVM is a SIC-POVM  \cite{Renes2004JMP}.
Despite the lack of a formal proof, it is widely believed that SIC-POVMs exist in any Hilbert space of finite dimension \cite{Zauner1999quantum}.
The SIC-POVM for qubits is also known as a tetrahedron measurement. It comprises four subnormalized projectors over pure states. 
In the Bloch sphere, each pure state is associated with a unit-length Bloch vector. 
The tips of the four Bloch vectors which characterize the SIC-POVM are the vertices of a regular tetrahedron \cite{Rehacek2004PRA}.

The determination of the quality of a given tomographic method is an important but complex problem.
The notion of optimality of a tomographic method depends on the assumptions made about the experimental setup (individual, collective, fixed or adaptative measurements), the reconstruction method employed (linear inversion, maximum likelihood estimation, etc), the particular quantifier of accuracy or efficiency (e.g., minimum squared error, maximum fidelity), and on how the average over the whole state of states is performed.
Under particular assumptions, it has been shown that SIC-POVMs are optimal among all minimal (those with the minimal number of outcomes) IC measurements \cite{Rehacek2004PRA,Scott2006JPA}, while MU measurements are optimal among all choices of IC projective measurements \cite{Zhu2014PRA}.
Moreover, several figures of merit and assumptions show that (optimal) MU measurements are more accurate than the corresponding SIC-POVMs \cite{Embacher2004,Roy2007,Zhu2014PRA,Rehacek2015PRA}.
Perhaps the most fundamental of these figures of merit is the quantum tomographic transfer function (the trace of the inverse of the Fisher matrix, averaged over pure states using the Haar measure), which gives the average optimal tomographic accuracy per sampling event for all unbiased state estimators, in the limit of a large number of sampling events \cite{Rehacek2015PRA}.

Numerous studies deal with the implementation of tomographic schemes in experiments involving photon polarization, including several schemes realizing this tetrahedron measurement \cite{Englert2005LP,Ling2006PRA,James2001PRA}. 
In contrast, fewer papers have been devoted to the tomographic reconstruction of double slit qubits. 
For example, a MUB approach was employed to measure the state of two double slit qubits \cite{Neves2007PRA,Lima2007JPCS}. 
In this case, detectors placed in the near and far-field of the slits, aided by double slit spatial filtering, allow the simultaneous measurement of the three Pauli operators of each qubit. 
It was subsequently discovered that spatial filtering is not essential, because Pauli operators can be measured using a lens and ``point" detectors in the image and focal planes. 
It was also recognized that, for a fixed detection-plane to slit-plane distance, the measurement of the interference pattern corresponds to a continuous POVM of a single qubit; therefore, the elements of the density operator can be obtained from the interference pattern  \cite{Taguchi2008PRA}.
A spatial light modulator, which can be used to control amplitudes and phases, was the key device to implement the minimal SIC tomography proposed in \cite{Rehacek2004PRA} to reconstruct double slit qubits \cite{Pimenta2010OE}.

Though optimal MU measurements are more accurate, SIC-POVMs are simpler.
Due to its simplicity,  we consider the implementation of the tetrahedron measurement of double slit qubits of light in this paper.
We show that a minimal SIC-POVM tomography of double slit qubits can be implemented by measures on a single plane, using only free propagation (without resorting to lenses, spatial light modulators, or other optical elements).

\section{Two-slit diffraction}
\label{sec:Sec2slit}
\begin{figure}[h!]
\begingroup
  \makeatletter
  \providecommand\color[2][]{%
    \GenericError{(gnuplot) \space\space\space\@spaces}{%
      Package color not loaded in conjunction with
      terminal option `colourtext'%
    }{See the gnuplot documentation for explanation.%
    }{Either use 'blacktext' in gnuplot or load the package
      color.sty in LaTeX.}%
    \renewcommand\color[2][]{}%
  }%
  \providecommand\includegraphics[2][]{%
    \GenericError{(gnuplot) \space\space\space\@spaces}{%
      Package graphicx or graphics not loaded%
    }{See the gnuplot documentation for explanation.%
    }{The gnuplot epslatex terminal needs graphicx.sty or graphics.sty.}%
    \renewcommand\includegraphics[2][]{}%
  }%
  \providecommand\rotatebox[2]{#2}%
  \@ifundefined{ifGPcolor}{%
    \newif\ifGPcolor
    \GPcolortrue
  }{}%
  \@ifundefined{ifGPblacktext}{%
    \newif\ifGPblacktext
    \GPblacktextfalse
  }{}%
  \let\gplgaddtomacro\g@addto@macro
  \gdef\gplbacktext{}%
  \gdef\gplfronttext{}%
  \makeatother
  \ifGPblacktext
    \def\colorrgb#1{}%
    \def\colorgray#1{}%
  \else
    \ifGPcolor
      \def\colorrgb#1{\color[rgb]{#1}}%
      \def\colorgray#1{\color[gray]{#1}}%
      \expandafter\def\csname LTw\endcsname{\color{white}}%
      \expandafter\def\csname LTb\endcsname{\color{black}}%
      \expandafter\def\csname LTa\endcsname{\color{black}}%
      \expandafter\def\csname LT0\endcsname{\color[rgb]{1,0,0}}%
      \expandafter\def\csname LT1\endcsname{\color[rgb]{0,1,0}}%
      \expandafter\def\csname LT2\endcsname{\color[rgb]{0,0,1}}%
      \expandafter\def\csname LT3\endcsname{\color[rgb]{1,0,1}}%
      \expandafter\def\csname LT4\endcsname{\color[rgb]{0,1,1}}%
      \expandafter\def\csname LT5\endcsname{\color[rgb]{1,1,0}}%
      \expandafter\def\csname LT6\endcsname{\color[rgb]{0,0,0}}%
      \expandafter\def\csname LT7\endcsname{\color[rgb]{1,0.3,0}}%
      \expandafter\def\csname LT8\endcsname{\color[rgb]{0.5,0.5,0.5}}%
    \else
      \def\colorrgb#1{\color{black}}%
      \def\colorgray#1{\color[gray]{#1}}%
      \expandafter\def\csname LTw\endcsname{\color{white}}%
      \expandafter\def\csname LTb\endcsname{\color{black}}%
      \expandafter\def\csname LTa\endcsname{\color{black}}%
      \expandafter\def\csname LT0\endcsname{\color{black}}%
      \expandafter\def\csname LT1\endcsname{\color{black}}%
      \expandafter\def\csname LT2\endcsname{\color{black}}%
      \expandafter\def\csname LT3\endcsname{\color{black}}%
      \expandafter\def\csname LT4\endcsname{\color{black}}%
      \expandafter\def\csname LT5\endcsname{\color{black}}%
      \expandafter\def\csname LT6\endcsname{\color{black}}%
      \expandafter\def\csname LT7\endcsname{\color{black}}%
      \expandafter\def\csname LT8\endcsname{\color{black}}%
    \fi
  \fi
  \setlength{\unitlength}{0.0500bp}%
  \begin{picture}(5472.00,2880.00)%
    \gplgaddtomacro\gplbacktext{%
      \csname LTb\endcsname%
      \put(783,1277){\makebox(0,0){\strut{}$2d$}}%
      \put(2443,2742){\makebox(0,0){\strut{}$z$}}%
      \put(1434,1521){\makebox(0,0){\strut{}$a$}}%
      \put(1434,1033){\makebox(0,0){\strut{}$a$}}%
      \put(3744,2791){\makebox(0,0){\strut{}$x$}}%
      \put(3508,1765){\makebox(0,0){\strut{}$x_{\mbox{\tiny 4}}$-}}%
      \put(3508,1440){\makebox(0,0){\strut{}$x_{\mbox{\tiny 3}}$-}}%
      \put(3508,1114){\makebox(0,0){\strut{}$x_{\mbox{\tiny 2}}$-}}%
      \put(3508,788){\makebox(0,0){\strut{}$x_{\mbox{\tiny 1}}$-}}%
    }%
    \gplgaddtomacro\gplfronttext{%
    }%
    \gplbacktext
    \put(0,0){\includegraphics{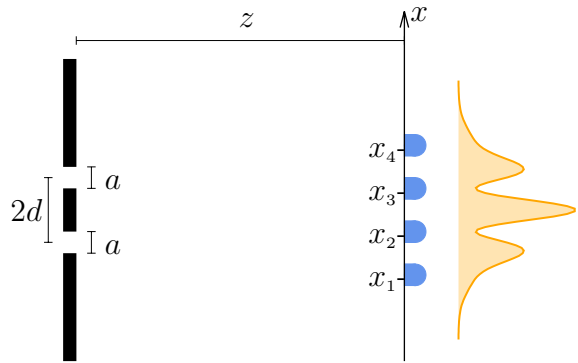}}%
    \gplfronttext
  \end{picture}%
\endgroup
\centering
\caption{(Color online) Double slit setup. Minimal tomography of double slit qubits requires photon measurements at four places in the detection plane. The interference pattern is shown on the right. }
\label{fig:1}
\end{figure}
We consider the double slit setup sketched in Fig. \ref{fig:1}, in which an electromagnetic plane wave propagates in the positive $z$ direction, towards a detection plane, after traversing a double slit screen.
The distance between the centers of the two slits, of width $a,$ is $2d.$
The midpoint between the slits is chosen as the origin of the coordinates. 
We assume quasimonochromatic waves, of frequency $\omega$ and wavenumber $k,$ in the paraxial approximation with a given polarization and a single spatial dimension $x$ on the slit and the detection planes. 
A simple mathematical description in terms of wave functions, recently used in the investigation of optical superresolution \cite{Tsang2016,Paur2016}, gives the same result as a second quantized treatment in which only the single photon subspace of the full Fock space is taken into account.

The light coming from the i-th slit is described by the wave function $\psi_i(x,z=0) = \braket{x|\psi_i(z=0)},$ $i=1,2.$
The paraxial wave equation
\begin{align*}
 \frac{\partial^2 \psi(x,z)}{\partial x^2} + 2i k \frac{\partial \psi(x,z) }{\partial z} =0.
\end{align*}
 connects the wave function at the slit plane with the corresponding wave function at the detection plane.
 It is convenient to switch to the dimensionless quantities $\xi={x}/{a},$ and $\zeta={z}/{z_0}.$
The paraxial wave equation can be simply written as $\partial_{\xi\xi} \psi +2 i \partial_{\zeta}\psi=0,$ where $z_0$ was chosen as $z_0=k a^2=2\pi a^2/\lambda.$ 
Here,  $\lambda$ is the wavelength of the  light that illuminates the double slit.
The general solution of the paraxial wave equation is a superposition of the elementary solutions  $e^{i\kappa\xi-i\kappa^2(\zeta/2)}$; that is, $\psi(\xi,\zeta) = \int d\kappa \tilde{\psi}(\kappa)e^{i\kappa\xi-i\kappa^2(\zeta/2)}$. 
Noticing that $\tilde{\psi}(\kappa)$ is the Fourier transform of $\psi(\xi^\prime,0)$, we have
\begin{align}\nonumber
 \psi(\xi,\zeta) & = \int \frac{d\kappa}{2\pi} \int d\xi^\prime \psi(\xi^\prime,0) e^{-i\kappa \xi^\prime} e^{i\kappa\xi-i\kappa^2\frac{\zeta}{2}}\\
  &= \int d\xi^\prime \frac{e^{i(\xi-\xi^\prime)^2/(2\zeta)}}{\sqrt{2\pi i \zeta}} \psi(\xi^\prime,0). \label{eq:evolution}
\end{align}
In the last step, the integral over  $\kappa$ was performed.

Under the assumptions made in this section, the most general state  of light just outside the slit screen is
\begin{align}
 \rho(\zeta=0) = \sum_{i,j=1,2} \rho_{ij}  \ket{\psi_i(\zeta=0)}\bra{\psi_j(\zeta=0)}.
\end{align}
A photon is detected at the (transversal) position $\xi$ on the detection plane with probability (density)
\begin{align*}
 p(\xi,\zeta) & =\braket{\xi|\rho(\zeta)|\xi}=
 \sum_{i,j=1,2} \rho_{ij}  \braket{\xi|\psi_i(\zeta)}\braket{\psi_j(\zeta)|\xi}\\
 & = \sum_{i,j=1,2} \psi_i(\xi,\zeta) \psi_j^*(\xi,\zeta) \rho_{ij}.
\end{align*}
In the following section, we show that a minimal SIC-POVM tomographic reconstruction of $\rho$, using photon detectors in a single plane, $\zeta=$ constant, is possible.

\section{Minimal double slit qubit tomography}
\label{sec:Solution}
\begin{figure}[htbp]
\begin{center}
\includegraphics[clip,scale=0.6]{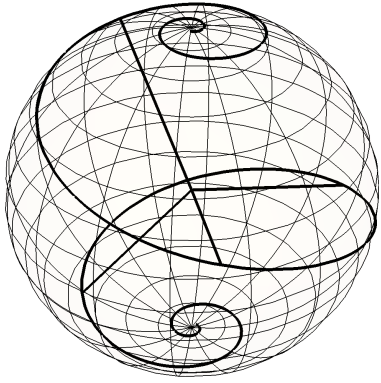}
\caption{(Color online)
As the transverse coordinate $\xi$ is swept from $-\infty$ to $\infty,$ the tip of the Bloch vector corresponding to the measurement state $\ket{\psi(\xi,\zeta_0)}$ goes from the north pole to the south pole.
The curve was calculated for Gaussian slits and $\zeta_0 \approx 3.4678.$ The straight lines are the Bloch vectors whose tips are the vertices of a tetrahedron.}
\label{fig:2}
\end{center}
\end{figure}
The photon detection probability density,  at the plane detection $\zeta$ and transversal position $\xi$, can be written in the suggestive form
\begin{align}
 p(\xi,\zeta) = I(\xi,\zeta)  \operatorname{Tr} \left( \hat{\Pi} (\xi,\zeta)  \rho(\zeta=0) \right),
\end{align}
where  $I(\xi,\zeta)= \sum_{k=1,2} |\psi_k(\xi,\zeta)|^2$ is the intensity envelope, and $\hat{\Pi} (\xi,\zeta)$, written in the basis $\{ \ket{\psi_1}, \ket{\psi_2} \}$,  is the projection operator
\begin{align*}
 \hat{\Pi} (\xi,\zeta) = \frac{1}{I(\xi,\zeta)}\begin{pmatrix} \psi_1(\xi,\zeta) \psi_1^*(\xi,\zeta) & \psi_1(\xi,\zeta) \psi_2^*(\xi,\zeta) \\ \psi_2(\xi,\zeta) \psi_1^*(\xi,\zeta) & \psi_2(\xi,\zeta) \psi_2^*(\xi,\zeta) \end{pmatrix}.
\end{align*}
Each projection operator projects over a pure state, $\varrho(\xi,\zeta) =\ket{\psi(\xi,\zeta)}\bra{\psi(\xi,\zeta)} $ $= \frac{1}{2}(\mathbb{I}+\boldsymbol{s}(\xi,\zeta)\cdot\boldsymbol{\sigma}),$ where $\mathbb{I}$ is the $2\times 2$ identity matrix, $\boldsymbol{\sigma}$ is a vector whose components are the three Pauli matrices, and $\boldsymbol{s}(\xi,\zeta)$ is the unit-length Bloch vector associated with the pure state $\varrho(\xi,\zeta).$
In a single plane, $\zeta=\zeta_0$ ($\zeta_0$ a constant),  the family of Bloch vectors $\boldsymbol{s}(\xi,\zeta_0)$ describes a curve over the Bloch sphere, as shown in Fig. \ref{fig:2}, as $\xi$ takes values on the real axis.

The photon detection probability density $p(\xi,\zeta)$, equal to the product of the intensity envelope $I(\xi,\zeta)$ and the population $0\leq \braket{\psi(\xi,\zeta)|\rho(0)|\psi(\xi,\zeta)}\leq 1,$ is bounded by the intensity envelope.
Therefore, $I(\xi,\zeta)$ is the maximum photon detection probability at a given point $\xi$ on the detection plane, as we measure over every possible state $\rho.$ 

The photon detection probability density $p(\xi,\zeta)$ depends on i) the four elements of the state $\rho(0)$ in the basis $\{\ket{\psi_1},\ket{\psi_2}\},$ ii) the two wave functions $\psi_1(\xi)$  and  $\psi_2(\xi),$ and iii) the evolution operator from the slit plane to the detection plane.
Besides the information on evolution, assumed by the usual state estimation schemes, we additionally assume complete knowledge of the wave functions $\psi_1(\xi)$  and  $\psi_2(\xi).$
Photon detection on a fixed detector plane $\zeta_0$ is described by the continuous POVM $\mathbb{I} = \int d\xi\, I(\xi,\zeta_0) \ket{\xi,\zeta_0}\bra{\xi,\zeta_0}.$
We want to use a minimal tomography, $\mathbb{M} = \sum_{i=1}^4 I(\xi_i,\zeta_0)\Delta\xi\,  \ket{\xi_i,\zeta_0}\bra{\xi_i,\zeta_0},$ which corresponds to four preselected positions of point detectors.
The spatial width of the photon detectors, $\Delta \xi,$ is chosen to be small enough to satisfy $\ket{\xi\pm\Delta\xi,\zeta_0}\approx \ket{\xi,\zeta_0}.$
In order to go from the continuous measurement to the minimal tomography, most of the information contained in the former measurement must be discarded.
Hence, the continuous POVM will generally give a better state estimation than the minimal tomography $\mathbb{M}.$ 
The minimal tomography, taking into account only the four point detectors, is modeled by $\mathbb{M} = \sum_{i=1}^4 P(\xi_i,\zeta_0) \ket{\xi_i,\zeta_0}\bra{\xi_i,\zeta_0},$ where the weigths $P(\xi_i,\zeta_0)$ are given by $I(\xi_i,\zeta_0)\Delta\xi/\Sigma(\zeta_0),$ and $\Sigma(\zeta_0) =  \sum_{i=1}^4 I(\xi_i,\zeta_0)\Delta\xi/2.$

It is well-known that the most accurate minimal POVM, $\sum_{i=1}^4 (1/2) \left( \mathbb{I} + \boldsymbol{s}_i\cdot \boldsymbol{\sigma}\right)/2 =\mathbb{I},$ projects over four pure states equally spaced on the Bloch sphere \cite{Rehacek2004PRA,Petz2012}; the tips of the corresponding Bloch vectors, $\boldsymbol{s}_i,\, i=1,2,3,4$, are the four vertices of a regular tetrahedron.
These Bloch vectors satisfy the inner product conditions, $\boldsymbol{s}_i \cdot \boldsymbol{s}_j = \frac{4}{3}\delta_{i,j} -\frac{1}{3},$ $i,j=1,2,3,4.$
If the family of Bloch vectors $\boldsymbol{s}(\xi,\zeta_0)$ contains four elements $\boldsymbol{s}(\xi_i,\zeta_0), i=1,2,3,4,$ which satisfy these conditions, a minimal tomographic reconstruction is possible if photon detectors are placed on the plane $\zeta=\zeta_0$, with transversal positions $\xi_i, i=1,2,3,4.$
Notice that the coefficients of the projectors are all equal.
Therefore, the intensity envelope, $I(\xi,\zeta_0),$ must be exactly the same for the four values of $\xi_i.$  

In this work, we focus on Gaussian slits, which can be thought of as an approximation to the rectangular slits used in actual experiments. 
Gaussian slits also describe the incidence of Gaussian beams on a biprism. 
Mathematically, Gaussian slits are modeled by the normalized wave functions
\begin{align}
 \psi_k(\xi,0) = \frac{1}{\sqrt[4]{\pi}} e^{-(\xi-(-1)^k \delta)^2/2}, \qquad k=1,2.
\end{align}
where $2\delta$ is the distance between the centers of the two slits, and the standard deviation of the slits is 1 ($a$ in the original variables). 
For $\zeta \geq 0$, we do have (for k=1,2)
\begin{align}
 \psi_k(\xi,\zeta) 
  = \int d\xi^\prime \frac{e^{i(\xi-\xi^\prime)^2/(2\zeta)}}{\sqrt{2\pi i \zeta}} \psi_k(\xi^\prime,0) =
  \frac{e^{-\frac{1-i\zeta}{2(1+\zeta^2)}(\xi-(-1)^k \delta)^2}}{\sqrt{\sqrt{\pi}(1+i\zeta)}}. 
\end{align}
Therefore, the  intensity envelope is
\begin{align*}
 I(\xi,\zeta;\delta) = \sum_{k=1,2} |\psi_k(\xi,\zeta)|^2 = \frac{2 e^{-\frac{\xi^2+\delta^2}{1+\zeta^2}}}{\sqrt{\pi(1+\zeta^2)}}  \cosh\left(\frac{2\delta\xi}{1+\zeta^2}\right).
\end{align*}
On the other hand, by setting $w=2\xi\delta/(1+\zeta^2)$, we can write the projectors as
\begin{align*}
\hat{\Pi}(\xi,\zeta) = \frac{1}{\cosh w} \begin{pmatrix} e^{-w} & e^{-i w \zeta}\\  e^{i w \zeta} & e^{ w} \end{pmatrix}
=\frac{1}{2} \begin{pmatrix} 1-\tanh(w) & \textrm{sech}(w) e^{-i w \zeta}\\ \textrm{sech}(w) e^{i w \zeta} & 1+\tanh(w) \end{pmatrix}.
\end{align*}
Taking into account that $\varrho(\xi,\zeta) = \frac{1}{2}(I+\boldsymbol{s}(\xi,\zeta)\cdot\boldsymbol{\sigma}),$ we obtain
\begin{align*}
\boldsymbol{s}(\xi,\zeta) = \textrm{sech}(w) \left(\cos(w \zeta), \sin(w \zeta), -\sinh(w) \right).
\end{align*}

To search four Bloch vectors $\boldsymbol{s}_i = \boldsymbol{s}(\xi_i,\zeta_0),$ $i=1,2,3,4,$ whose tips are the vertices of a tetrahedron, we define the positive function 
$f(\boldsymbol{s}_1,\boldsymbol{s}_2,\boldsymbol{s}_3,\boldsymbol{s}_4) = \sum_{i>j} (\boldsymbol{s}_i\cdot\boldsymbol{s}_j+1/3)^2,$ which vanishes only when the tetrahedron condition is met.
Rough local minima of $f$, obtained with an algorithm that uses random values of $w$ and $\zeta$, are fed to a gradient algorithm, which provides a more accurate approximation.
Some solutions, corresponding to the smallest values of $\zeta$ found, are listed in Table \ref{table:0}.
Notice the existence of solutions which are not symmetric under  reflection, $\xi\to -\xi.$
 \begin{table}[htbp]
\caption{Tetrahedron tomography: some numerical solutions}
\begin{center}
\begin{tabular}{ccccc} \hline
 $\zeta$ & $w_1$  & $w_2$ & $w_3$ & $w_3$ \\
\hline
3.4678 & -1.0287 & -0.268044 & 0.268044 & 1.0287  \\
6.08028 &  -0.943335 & -0.367661 & 0.367661 & 0.943335 \\
7.70501 &  -1.89747 & 0.0782496& 0.332645 &0.628523\\
8.5243 & -1.30348 & -0.153589 &0.363175 & 0.805324 \\
8.55362 &  -1.124 &  -0.111174 & 0.111174 &1.124\\
10.6561 & -0.738203 & -0.487839 & 0.272693 & 1.140046\\
\end{tabular}
\end{center}
\label{table:0}
\end{table}%

The four Bloch vectors, for the smallest $\zeta$ solution, are 
\begin{align*}
\boldsymbol{s}_1 &=\left(-1/\sqrt{3}, -0.261804, 0.773386\right), \\ \boldsymbol{s}_2&=\left(1/\sqrt{3}, 0.773386, 0.261804\right),\\
\boldsymbol{s}_3& =\left(1/\sqrt{3}, -0.773386, -0.261804\right), \\ \boldsymbol{s}_4&=\left(-1/\sqrt{3},  0.261804, -0.773386\right).
\end{align*}
They are drawn in Fig. \ref{fig:2}, along with the set of projectors obtained by varying the transversal position of the detector.
It is easy to see that the $s_x=\pm 1/\sqrt{3}$ for symmetric solutions. Indeed, since $\boldsymbol{s}(\xi,\zeta)$ is a unit-length vector and $\boldsymbol{s}(-\xi,\zeta)=(s_x(\xi,\zeta),-s_y(\xi,\zeta),-s_z(\xi,\zeta)),$ we have  $\boldsymbol{s}(-\xi,\zeta)\cdot \boldsymbol{s}(\xi,\zeta) = 2s_x^2(\xi,\zeta)-1.$ Finally, when  $\boldsymbol{s}(\xi,\zeta)$  and $\boldsymbol{s}(-\xi,\zeta)$ belong to a tetrahedron, its inner product is $-1/3=2s_x^2(\xi,\zeta)-1$. The desired result follows from this equality.

\begin{figure}[htb]
\begingroup
  \makeatletter
  \providecommand\color[2][]{%
    \GenericError{(gnuplot) \space\space\space\@spaces}{%
      Package color not loaded in conjunction with
      terminal option `colourtext'%
    }{See the gnuplot documentation for explanation.%
    }{Either use 'blacktext' in gnuplot or load the package
      color.sty in LaTeX.}%
    \renewcommand\color[2][]{}%
  }%
  \providecommand\includegraphics[2][]{%
    \GenericError{(gnuplot) \space\space\space\@spaces}{%
      Package graphicx or graphics not loaded%
    }{See the gnuplot documentation for explanation.%
    }{The gnuplot epslatex terminal needs graphicx.sty or graphics.sty.}%
    \renewcommand\includegraphics[2][]{}%
  }%
  \providecommand\rotatebox[2]{#2}%
  \@ifundefined{ifGPcolor}{%
    \newif\ifGPcolor
    \GPcolortrue
  }{}%
  \@ifundefined{ifGPblacktext}{%
    \newif\ifGPblacktext
    \GPblacktextfalse
  }{}%
  \let\gplgaddtomacro\g@addto@macro
  \gdef\gplbacktext{}%
  \gdef\gplfronttext{}%
  \makeatother
  \ifGPblacktext
    \def\colorrgb#1{}%
    \def\colorgray#1{}%
  \else
    \ifGPcolor
      \def\colorrgb#1{\color[rgb]{#1}}%
      \def\colorgray#1{\color[gray]{#1}}%
      \expandafter\def\csname LTw\endcsname{\color{white}}%
      \expandafter\def\csname LTb\endcsname{\color{black}}%
      \expandafter\def\csname LTa\endcsname{\color{black}}%
      \expandafter\def\csname LT0\endcsname{\color[rgb]{1,0,0}}%
      \expandafter\def\csname LT1\endcsname{\color[rgb]{0,1,0}}%
      \expandafter\def\csname LT2\endcsname{\color[rgb]{0,0,1}}%
      \expandafter\def\csname LT3\endcsname{\color[rgb]{1,0,1}}%
      \expandafter\def\csname LT4\endcsname{\color[rgb]{0,1,1}}%
      \expandafter\def\csname LT5\endcsname{\color[rgb]{1,1,0}}%
      \expandafter\def\csname LT6\endcsname{\color[rgb]{0,0,0}}%
      \expandafter\def\csname LT7\endcsname{\color[rgb]{1,0.3,0}}%
      \expandafter\def\csname LT8\endcsname{\color[rgb]{0.5,0.5,0.5}}%
    \else
      \def\colorrgb#1{\color{black}}%
      \def\colorgray#1{\color[gray]{#1}}%
      \expandafter\def\csname LTw\endcsname{\color{white}}%
      \expandafter\def\csname LTb\endcsname{\color{black}}%
      \expandafter\def\csname LTa\endcsname{\color{black}}%
      \expandafter\def\csname LT0\endcsname{\color{black}}%
      \expandafter\def\csname LT1\endcsname{\color{black}}%
      \expandafter\def\csname LT2\endcsname{\color{black}}%
      \expandafter\def\csname LT3\endcsname{\color{black}}%
      \expandafter\def\csname LT4\endcsname{\color{black}}%
      \expandafter\def\csname LT5\endcsname{\color{black}}%
      \expandafter\def\csname LT6\endcsname{\color{black}}%
      \expandafter\def\csname LT7\endcsname{\color{black}}%
      \expandafter\def\csname LT8\endcsname{\color{black}}%
    \fi
  \fi
  \setlength{\unitlength}{0.0500bp}%
  \begin{picture}(5328.00,3888.00)%
    \gplgaddtomacro\gplbacktext{%
      \csname LTb\endcsname%
      \put(946,704){\makebox(0,0)[r]{\strut{}$0$}}%
      \put(946,1580){\makebox(0,0)[r]{\strut{}$0.15$}}%
      \put(946,2455){\makebox(0,0)[r]{\strut{}$0.3$}}%
      \put(946,3331){\makebox(0,0)[r]{\strut{}$0.45$}}%
      \put(1078,484){\makebox(0,0){\strut{}$-12$}}%
      \put(1720,484){\makebox(0,0){\strut{}$-8$}}%
      \put(2362,484){\makebox(0,0){\strut{}$-4$}}%
      \put(3005,484){\makebox(0,0){\strut{}$0$}}%
      \put(3647,484){\makebox(0,0){\strut{}$4$}}%
      \put(4289,484){\makebox(0,0){\strut{}$8$}}%
      \put(4931,484){\makebox(0,0){\strut{}$12$}}%
      \put(176,2163){\rotatebox{-270}{\makebox(0,0){\strut{}$I(\xi)$}}}%
      \put(3136,154){\makebox(0,0){\strut{}$\xi$}}%
    }%
    \gplgaddtomacro\gplfronttext{%
      \csname LTb\endcsname%
      \put(4358,3349){\makebox(0,0)[r]{\strut{}$\delta=2.76$}}%
      \csname LTb\endcsname%
      \put(4358,3151){\makebox(0,0)[r]{\strut{}$\delta=1.50$}}%
      \csname LTb\endcsname%
      \put(4358,2953){\makebox(0,0)[r]{\strut{}$\delta=3.50$}}%
    }%
    \gplbacktext
    \put(0,0){\includegraphics{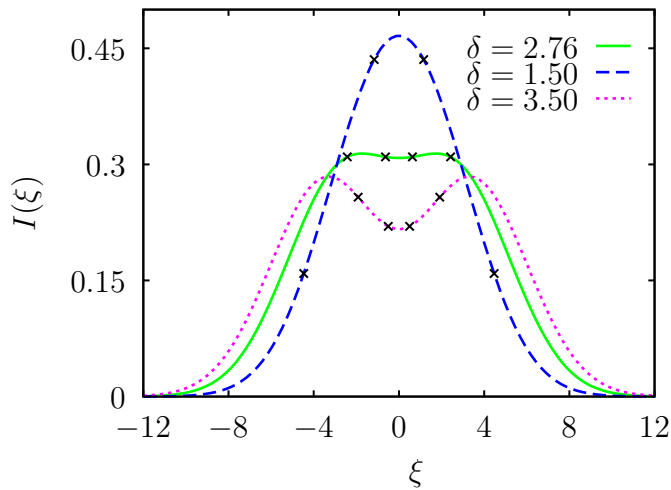}}%
    \gplfronttext
  \end{picture}%
\endgroup
\centering
\caption{(Color online) Intensity envelope for three different distances between the slits. The values of the transverse coordinate, where the photon detectors must be placed for a tetrahedron measurement, are marked with crosses. The tetrahedron measurement becomes unbiased  when $\delta \approx 2.76$.  }
\label{fig:3}
\end{figure}

\begin{figure}[hbt]
\begingroup
  \makeatletter
  \providecommand\color[2][]{%
    \GenericError{(gnuplot) \space\space\space\@spaces}{%
      Package color not loaded in conjunction with
      terminal option `colourtext'%
    }{See the gnuplot documentation for explanation.%
    }{Either use 'blacktext' in gnuplot or load the package
      color.sty in LaTeX.}%
    \renewcommand\color[2][]{}%
  }%
  \providecommand\includegraphics[2][]{%
    \GenericError{(gnuplot) \space\space\space\@spaces}{%
      Package graphicx or graphics not loaded%
    }{See the gnuplot documentation for explanation.%
    }{The gnuplot epslatex terminal needs graphicx.sty or graphics.sty.}%
    \renewcommand\includegraphics[2][]{}%
  }%
  \providecommand\rotatebox[2]{#2}%
  \@ifundefined{ifGPcolor}{%
    \newif\ifGPcolor
    \GPcolortrue
  }{}%
  \@ifundefined{ifGPblacktext}{%
    \newif\ifGPblacktext
    \GPblacktextfalse
  }{}%
  \let\gplgaddtomacro\g@addto@macro
  \gdef\gplbacktext{}%
  \gdef\gplfronttext{}%
  \makeatother
  \ifGPblacktext
    \def\colorrgb#1{}%
    \def\colorgray#1{}%
  \else
    \ifGPcolor
      \def\colorrgb#1{\color[rgb]{#1}}%
      \def\colorgray#1{\color[gray]{#1}}%
      \expandafter\def\csname LTw\endcsname{\color{white}}%
      \expandafter\def\csname LTb\endcsname{\color{black}}%
      \expandafter\def\csname LTa\endcsname{\color{black}}%
      \expandafter\def\csname LT0\endcsname{\color[rgb]{1,0,0}}%
      \expandafter\def\csname LT1\endcsname{\color[rgb]{0,1,0}}%
      \expandafter\def\csname LT2\endcsname{\color[rgb]{0,0,1}}%
      \expandafter\def\csname LT3\endcsname{\color[rgb]{1,0,1}}%
      \expandafter\def\csname LT4\endcsname{\color[rgb]{0,1,1}}%
      \expandafter\def\csname LT5\endcsname{\color[rgb]{1,1,0}}%
      \expandafter\def\csname LT6\endcsname{\color[rgb]{0,0,0}}%
      \expandafter\def\csname LT7\endcsname{\color[rgb]{1,0.3,0}}%
      \expandafter\def\csname LT8\endcsname{\color[rgb]{0.5,0.5,0.5}}%
    \else
      \def\colorrgb#1{\color{black}}%
      \def\colorgray#1{\color[gray]{#1}}%
      \expandafter\def\csname LTw\endcsname{\color{white}}%
      \expandafter\def\csname LTb\endcsname{\color{black}}%
      \expandafter\def\csname LTa\endcsname{\color{black}}%
      \expandafter\def\csname LT0\endcsname{\color{black}}%
      \expandafter\def\csname LT1\endcsname{\color{black}}%
      \expandafter\def\csname LT2\endcsname{\color{black}}%
      \expandafter\def\csname LT3\endcsname{\color{black}}%
      \expandafter\def\csname LT4\endcsname{\color{black}}%
      \expandafter\def\csname LT5\endcsname{\color{black}}%
      \expandafter\def\csname LT6\endcsname{\color{black}}%
      \expandafter\def\csname LT7\endcsname{\color{black}}%
      \expandafter\def\csname LT8\endcsname{\color{black}}%
    \fi
  \fi
  \setlength{\unitlength}{0.0500bp}%
  \begin{picture}(5328.00,3888.00)%
    \gplgaddtomacro\gplbacktext{%
      \csname LTb\endcsname%
      \put(814,704){\makebox(0,0)[r]{\strut{}$0$}}%
      \put(814,1677){\makebox(0,0)[r]{\strut{}$0.2$}}%
      \put(814,2650){\makebox(0,0)[r]{\strut{}$0.4$}}%
      \put(814,3623){\makebox(0,0)[r]{\strut{}$0.6$}}%
      \put(946,484){\makebox(0,0){\strut{}$-6$}}%
      \put(1610,484){\makebox(0,0){\strut{}$-4$}}%
      \put(2274,484){\makebox(0,0){\strut{}$-2$}}%
      \put(2939,484){\makebox(0,0){\strut{}$0$}}%
      \put(3603,484){\makebox(0,0){\strut{}$2$}}%
      \put(4267,484){\makebox(0,0){\strut{}$4$}}%
      \put(4931,484){\makebox(0,0){\strut{}$6$}}%
      \put(176,2163){\rotatebox{-270}{\makebox(0,0){\strut{}$\psi(\xi)$}}}%
      \put(2938,154){\makebox(0,0){\strut{}$\xi$}}%
    }%
    \gplgaddtomacro\gplfronttext{%
      \csname LTb\endcsname%
      \put(4076,3270){\makebox(0,0)[r]{\strut{}$\psi_1$}}%
      \csname LTb\endcsname%
      \put(4076,3050){\makebox(0,0)[r]{\strut{}$\psi_2$}}%
    }%
    \gplbacktext
    \put(0,0){\includegraphics{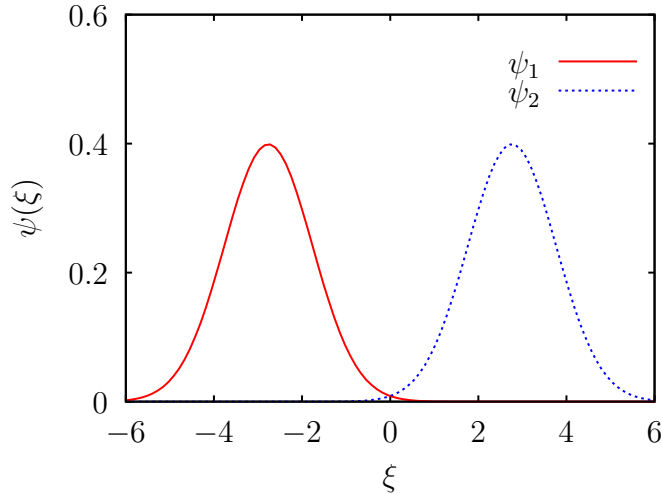}}%
    \gplfronttext
  \end{picture}%
\endgroup
\centering
\caption{(Color online) Mode profiles for optimal tomography display negligible overlap at $\zeta=0$.}
\label{fig:4}
\end{figure}

The numerical solutions that we have found, described by $\mathbb{M} = \sum_{i=1}^4\mathbb{P}_i= \sum_{i=1}^4 P(\xi_i,\zeta_0) \ket{\xi_i,\zeta_0}\bra{\xi_i,\zeta_0},$ are not true solutions, in the sense that the subnormalized projectors $\mathbb{P}_i$ do not add to the identity matrix.
This problem originates on the fact that the weights $P(\xi_i,\zeta_0),$ which are proportional to the intensity envelope $I(\xi_i,\zeta_0),$ are not equal.
To deal with this problem, one can artificially ``balance'' the actual counts, multiplying them by an appropriate constant, as have been done in some experimental realizations of minimal tomography.
A better alternative would be to examine the behavior of the intensity envelope as a function of the transverse coordinate and the slit separation (see Fig. \ref{fig:3}), with the aim to find solutions intrinsically balanced.
When we write the intensity envelope $I(\xi,\zeta;\delta)$ as $2\exp({-\frac{(1+\zeta^2)w^2}{4\delta^2}}) \exp(-\frac{\delta^2}{1+\zeta^2})  \cosh (w)/\sqrt{\pi(1+\zeta^2)}$, we see that the symmetric solutions are balanced when the distance between the slits is chosen as
\begin{align}\label{Eq:delta}
 \delta = \frac{1}{2} \sqrt{\frac{(1+\zeta^2)(w_1^2-w_2^2)}{\ln \cosh w_1- \ln \cosh w_2}},
\end{align}
where $w_1<w_2<0.$
Indeed, when $\delta$ is chosen according to \eqref{Eq:delta}, the value of the intensity envelope is the same at the four photon detector positions.
In the case of the solution at $\zeta \approx 3.4678$, this tomography is  a POVM for $\delta\approx 2.76444.$
In Fig. \ref{fig:3}, we have plotted the  intensity envelope as a function of the dimensionless transverse coordinate $\xi,$ for several values of the dimensionless distance between the centers of the slits, $\delta.$
The curves have been rescaled in such a way that, for $\zeta=0,$ its maximum value would be unity.
Moreover, as a final check, the two slits can be seen to be non-overlapping, by plotting $\psi_1(\xi,\zeta=0,\delta)$ and $\psi_2(\xi,\zeta=0,\delta)$ (Fig. \ref{fig:4}).
In fact, $\braket{\psi_1(\xi,\zeta;\delta)|\psi_2(\xi,\zeta;\delta)}=e^{-\delta^2}$ is of the order of $10^{-4}$ for  $\delta\approx 2.76.$ 

\section{Final remarks}

In this paper, we consider the implementation of the SIC tetrahedron tomography for double slit qubits of light using measurements performed on a single plane.
We have found that $\lambda$, the wavelength of the photons used in the experiment, and $a$, the width of the slits, can be used to determine the geometry of the experimental setup which implements the optimal state estimation, and does not require to artificially balance the measured probabilities.
The particular details of the solution depend on the states which are propagated from the double slit screen.
Assuming plane waves and Gaussian slits, the optimal geometry, for detectors placed nearest to the double slit plane, features a distance between the centers of the slits equal to 5.53 $a$.
The distance between the slit-plane and the detection-plane is $3.47\times 2\pi a^2/\lambda$.
The detectors must be placed at the transversal positions $x_1=-x_4,$ $x_2=-x_3,$ $x_3$ and $x_4$, measured from the midpoint between the slits (but on the detection plane).
In Table \ref{table:1}, we give some typical values of the geometry, by using wavelengths and slit widths reported in the literature.
 \begin{table}[htbp]
\caption{Optimal geometry for nearest SIC tetrahedron tomography of double slit qubits}
\begin{center}
\begin{tabular}{cccccc} \hline
 $\lambda$ & $a$  &$2d_{opt}$ & $z_0$ & $x_3$ & $x_4$\\
\hline
650 nm& 100 $\mu$m & 553 $\mu$m & 33.53 cm & 63 $\mu$m & 242 $\mu$m  \\
780 nm &  62.5 $\mu$m & 346 $\mu$m & 10.9 cm &39 $\mu$m & 151 $\mu$m \\
826 nm &  60 $\mu$m & 332 $\mu$m & 9.5 cm &38 $\mu$m & 145 $\mu$m\\
810 nm &  40 $\mu$m & 221 $\mu$m & 3.5 cm &25 $\mu$m & 97 $\mu$m\\
\end{tabular}
\end{center}
\label{table:1}
\end{table}%

The results of this work can be applied not only to the classical version of this setup, but also to spatial qubits of matter.
Moreover, the tomographic reconstruction of two double slit qubits can be carried out by joint measurements of two single qubit SIC tetrahedron projectors.
An approach similar to the one used in this work makes it possible to investigate the implementation of the SIC tomography for d-slit qudits of light, using measurements performed on a minimum number of planes.

\section*{Acknowledgements}

We thank Prof. M. C. Nemes for suggesting this problem, Dr. Breno Marques for stimulating discussions, and an anonymous referee for his/her constructive input.
This research did not receive any specific grant from funding agencies in the public, commercial, or not-for-profit sectors.


\end{document}